\documentclass[iop,numberedappendix,apj]{emulateapj}

\usepackage{color}
\usepackage{ulem}
\usepackage{amsmath}

\def\bE{{\mathbf E}}
\def\bB{{\mathbf B}}
\def\bj{{\mathbf j}}
\def\bv{{\mathbf v}}
\def\br{{\mathbf r}}
\def\bOm{{\mathbf \Omega}}
\def\rhoGJ{\rho_{\rm GJ}}
\def\beq{\begin{equation}}
\def\eeq{\end{equation}}
\def\Eq{Equation}
\def\gthr{\gamma_{\rm thr}}
  \def\tev{t_{\rm ohm}}
\def\tA{t_{\rm A}}
\def\gDL{\gamma_{\rm DL}}
\def\fmax{f_{\max}}
\def\tA{t_{\rm A}}
\def\Etw{E_{\rm twist}}
\def\rout{r_{\rm out}}
\def\ttw{t_{\rm shear}}
\def\Rs{R_\star}
\def\Bs{B_\star}

\def\lgap{\ell_{\rm gap}}
\def\Egap{E_{\rm gap}}
\def\Phigap{\Phi_{\rm gap}}
\def\Phithr{\Phi_{\rm thr}}
\def\M{{\cal M}}

\newbox\grsign \setbox\grsign=\hbox{$>$} \newdimen\grdimen \grdimen=\ht\grsign
\newbox\simlessbox \newbox\simgreatbox \newbox\simpropbox
\setbox\simgreatbox=\hbox{\raise.5ex\hbox{$>$}\llap
     {\lower.5ex\hbox{$\sim$}}}\ht1=\grdimen\dp1=0pt
\setbox\simlessbox=\hbox{\raise.5ex\hbox{$<$}\llap
     {\lower.5ex\hbox{$\sim$}}}\ht2=\grdimen\dp2=0pt
\setbox\simpropbox=\hbox{\raise.5ex\hbox{$\propto$}\llap
     {\lower.5ex\hbox{$\sim$}}}\ht2=\grdimen\dp2=0pt
\def\simgt{\mathrel{\copy\simgreatbox}}
\def\simlt{\mathrel{\copy\simlessbox}}

\begin{document}

\title{Particle-in-cell simulations of the twisted magnetospheres of magnetars}

\author{Alexander Y. Chen, Andrei M. Beloborodov}
\affil{Physics Department and Columbia Astrophysics Laboratory,
  Columbia University, 538  West 120th Street New York, NY 10027
}

\begin{abstract}
The magnetospheres of magnetars are believed to be filled with electron-positron
plasma generated by electric discharge. We present a first direct numerical experiment
showing how the plasma is created in an axisymmetric closed magnetosphere.
The $e^\pm$ discharge occurs  in response to twisting of the magnetic field lines
by a shear deformation of the magnetar surface, which launches electric currents
into the magnetosphere. The simulation shows the formation of
an electric ``gap'' with unscreened electric field ($\bE\cdot \bB\neq 0$) that continually
accelerates particles along the magnetic field lines and sustains pair creation.
The accelerating voltage is self-regulated to the threshold of the $e^\pm$ discharge.
It controls the rate of energy release and the lifetime of the magnetic twist. The simulation
follows the global evolution of the twisted magnetosphere over a long time and
demonstrates its gradual resistive untwisting. A vacuum cavity forms near the star
and expands, gradually erasing magnetospheric electric currents $j$. The active j-bundle
shrinks with time and its footprints form shrinking hot spots on the magnetar surface
bombarded by the created particles.
\bigskip
\end{abstract}

\bigskip
\section{Introduction}
\label{sec:intro}

Magnetars are neutron stars with ultrastrong magnetic fields ($B \gtrsim
10^{14}\,\mathrm{G}$) that display  strong activity fed by dissipation of magnetic energy
\citetext{see e.g. \citealp{mereghetti_2008}; \citealp{2015RPPh...78k6901T} for reviews}.
They produce strong outbursts and flares as well as bright persistent emission
with a prominent hard X-ray component extending above 100~keV.
These activities are associated with strong
deformations of the external magnetosphere of the neutron star, resembling the activity
of the solar corona
\citep[e.g.][]{1995MNRAS.275..255T}.
The magnetosphere is anchored in the solid crust of the star and its deformation
is caused by crustal shear motions driven by ultrastrong internal magnetic
stresses.

The speed of the surface motions is poorly known. Recent work suggests that
the crust yields to internal stresses through an instability launching a
thermoplastic wave
\citep{2014ApJ...794L..24B}
or a Hall-mediated avalanche
\citep{2016arXiv160604895L}
In both cases the motion is plastic and should occur on a timescale much longer
than the Alfv\'en crossing timescale ($10-100\,\mathrm{ms}$).
It is expected to be fast enough to efficiently twist the external magnetosphere.

The surface shear motion launches Alfv\'en waves along the magnetic field lines and
generates magnetospheric twist $\nabla\times \mathbf{B} \neq 0$
\citetext{\citealp{2002ApJ...574..332T}; \citealp{2013ApJ...774...92P},
  hereafter PBH13}.
Plasma is required to supply the current $\mathbf{j} = (c/4\pi)\nabla\times \mathbf{B}$.
\citet[hereafter BT07]{beloborodov_corona_2007}
found that plasma must be mainly supplied through $e^\pm$ discharge in the
magnetosphere rather than through extraction of charges from the star.
They performed simplified one-dimensional (1D) simulations of the discharge.
In the simulations, the magnetosphere was replaced by
a fixed, uniform field $\bB(x)$ connecting anode and cathode ---
metallic plates at $x_A$ and $x_C$. The fixed $\nabla\times\bB$ in this
setup turns out equivalent to imposing an electric current through the plates into
the computational box.
When pair creation was not allowed, the system quickly relaxed to a global
``double layer'' configuration, with surface charges of the opposite sign induced on the plates.
The electric field between them gave a huge voltage $\Phi_e$ accelerating
particles to ultra-high energies. When pair creation process was included in the simulation,
the voltage dropped to a much lower value, just sufficient to sustain pair creation,
and the current was supported through continual $e^\pm$ discharge.
BT07 concluded that pair creation must be responsible for screening
electric fields and regulating the magnetospheric activity of magnetars.

The simplified 1D model cannot, however, give a compete picture of the
magnetospheric activity, for a few reasons. It does not show how
$\nabla\times\bB$ is imparted in the first place, as the 1D model does not
support Alfv\'en waves. The exclusion of this important degree of freedom may
also put in question the double layer formation in the absence of pair creation,
the necessity of the onset of pair creation, and the self-regulation of the
discharge voltage seen in the 1D model. Note also that the electric field in the
1D (slab) geometry does not decrease with distance from the charge, and hence
one cannot see a realistic distribution of the accelerating electric field along
the magnetospheric field lines. Finally, the 1D model offers no way to follow
the gradual resistive ``untwisting'' of the magnetosphere --- its global
evolution as a result of ohmic dissipation of the twist energy. The expected
evolution must occur on the resistive timescale of months to years (regulated by
voltage $\Phi_e$) and can be tested against observations.

An axisymmetric electrodynamic model of a resistively untwisting magnetosphere
was developed by
\citet[hereafter B09]{2009ApJ...703.1044B}.
This model assumed
that a given fixed voltage $\Phi_e$ is sustained on current-carrying field lines,
without calculating the discharge that regulates $\Phi_e$.
A surprising result was
the formation of two distinct regions in the untwisting magnetosphere, with a
sharp boundary between them, --- a ``cavity'' ($j=0$) and a ``j-bundle.'' In
essence, the untwisting process was found to be the growth of the cavity,
erasing the currents in the j-bundle. A curious immediate implication was the
prediction of shrinking hot spots on magnetars --- the footprints of the
shrinking j-bundle, where the stellar surface is heated by bombardment of
  accelerated particles. Shrinking hot spots have been observed in seven
objects by now
\citetext{see data compilation in \citealp{2016arXiv160509077B}}.
All of these objects
belong to the class of ``transient magnetars'' that show a sudden outburst and then
gradually decay back to the quiescent state of low luminosity.
A key parameter governing the j-bundle evolution is its poorly known voltage $\Phi_e$,
which depends on how the $e^\pm$ discharge is self-organized and
may be different on different magnetospheric field lines.

The goal of the present paper is to overcome the limitations of the 1D discharge
model and perform a first self-consistent calculation of the $e^\pm$ discharge
in an axisymmetric twisted magnetosphere. The process can be simulated from
first-principles using a full kinetic description of the magnetospheric plasma
as a large number of charged particles moving in the self-consistent collective
electromagnetic field. Such a direct numerical experiment will show how the
twist and the electric current are created in the magnetosphere in response to
crustal shear, and will follow the ensuing dissipative evolution of the twist.

The self-organization of the $e^\pm$ discharge should determine where the
particles are created and accelerated. Should this occur near the footpoints of
the magnetospheric field line or near its apex? Will the acceleration region be
steady or move around? Answers to these questions may have important
implications for nonthermal emission from magnetars. The voltage drop along the
twisted field lines will control the dissipated power which feeds the observed
emission. We expect to see how particles are accelerated in the current-carrying
magnetic loop and rain down on the stellar surface to create hotspots. Finally,
the established discharge voltage will determine the life-time of the magnetic
twist and the pattern of its evolution.

A suitable technique for such direct numerical experiments is the particle-in-cell
(PIC) method, with pair creation implemented. This method has been
successfully applied to the old problem of rotation-powered pulsars
\citetext{\citealp{chen_pulsar_2014}, hereafter CB14; \citealp{2015ApJ...801L..19P}; \citealp{2015MNRAS.449.2759B}; \citealp{2016MNRAS.457.2401C}}
The magnetar problem is different in important ways and in some ways
easier to study using a global PIC simulation, as will be described below.

The paper is organized as follows. In Section~\ref{sec:theory} we describe the theory of
twisted magnetospheres in axisymmetric geometry, revisit the double-layer
configuration (in the absence of pair creation), describe
the mechanism of pair creation and basic electrodynamics of untwisting.
This will be useful for understanding the simulation results and also introduces
notation used in the paper.
Section~\ref{sec:setup} presents the setup of our numerical experiments.
Section~\ref{sec:result} describes the results and their implications.
Finally in Section~\ref{sec:discussion} we summarize our conclusions and provide an outlook for future studies.


\section{Sustaining Currents in the Twisted Magnetosphere}
\label{sec:theory}

Let us consider a dipole magnetic field around the star,
and assume that its footpoints on the star are sheared in the azimuthal direction
about the magnetic axis. In this case,
the implanted twist is axisymmetric and its amplitude $\psi$ is
simply
given by the azimuthal
angle between the two footpoints of the magnetic field line.
It is convenient to use spherical coordinates $r,\theta,\phi$ with the polar axis being
the axis of symmetry. The magnetospheric twist implies a toroidal component of
the magnetic field $B_\phi\neq 0$, and the twist amplitude on a given magnetic field
line is related to $B_\phi$ by
\begin{equation}
  \label{eq:twist-angle}
  \psi = \int_p^q \frac{B_{\phi}}{B\,r\sin\theta}d \ell,
\end{equation}
where the integral is taken along the field line, and $p$, $q$ are the two
footpoints where the field line is anchored to the surface. As long as the
implanted twist $\psi$ is smaller than unity, the poloidal magnetic field
remains close to dipolar, and the deformation can be thought of as simply adding
a toroidal component $B_{\phi}$ without changing the poloidal dipole component
(B09). This induces $\nabla\times \mathbf{B}$ in the dipolar configuration that
was originally curl-free. It must be sustained by an electric current in the
magnetosphere, $\bj$. The magnetic energy strongly dominates over the plasma
energy, and hence the currents must be nearly force-free, $\bj\times\bB=0$, i.e.
flowing along the magnetic field lines.

The origin of the plasma that could carry the current is a non-trivial issue.
The star can have a gaseous atmosphere, however for the typical surface
temperature $kT<1$~keV the atmosphere scale-height is tiny (centimeters),
because of the strong gravity of the neutron star. The atmosphere does not provide
enough plasma to conduct currents at large altitudes $r\sim \Rs$, where
$\Rs=10-13$~km is the neutron star radius.

Spinning of the neutron star and its magnetosphere with velocity
$\bv_{\rm rot}=\bOm\times\br$ implies a ``co-rotation'' electric field
$\bE=-\bv_{\rm rot}\times\bB/c$ and requires charge density
$\rho_\mathrm{GJ}=\nabla\cdot\bE/4\pi=-\bOm\cdot \mathbf{B}/2\pi c$
\citep{goldreich_pulsar_1969}.
Magnetars are slow rotators, $\Omega\sim 1$~Hz,
and their $\rhoGJ$ is small. The currents demanded by the twisted magnetosphere
are typically much stronger than $c\rhoGJ$.

The magnetosphere must make a
special effort to avoid charge starvation and create sufficiently dense plasma
to conduct the current $\bj$ demanded by the twist.
It achieves this by inducing an electric field $E_\parallel$ (parallel to the
magnetic field lines) that can
accelerate particles
and trigger pair creation.
This implies a finite voltage in the magnetospheric electric circuit
and a finite rate of ohmic dissipation.

\subsection{Voltage without pair discharge}
\label{sec:v-no-pair}

In the absence of pair creation,
the star is the only available source of
magnetospheric
plasma. The lack of charges leads to induction of an electric field with
a component parallel to the magnetic field, which can pull out charges from
the star and accelerate them.
Then the
electric circuit is expected to relax to a static configuration
similar to the relativistic double layer derived by
\citet{1982Ap&SS..87...21C}
and observed in the 1D plasma simulations of BT07.
It sustains the opposite surface charges at the two footpoints of the magnetic loop
where the lifted particles still move slowly, $v\ll c$, and create a large charge density
$\rho\sim j/v$.

The high charge density near the footpoints generates $E_\parallel$ according to
the Gauss law, and $E_\parallel$ accelerates the flow on the plasma timescale
$\omega_p^{-1}=(m_e/4\pi e\rho)^{1/2}$. The flow density $\rho$ is reduced to
its minimum where its velocity approaches $c$. As a result, the characteristic
thickness of the surface charge layer is the plasma skin depth
$\lambda_p=c/\omega_p$ evaluated for the plasma density $\rho\sim j/c$.

The surface charge $\Sigma\sim (j/c)\lambda_p$ generates the self-consistent
electric field that lifts and accelerates particles from the footpoint,
\begin{equation}
  \label{eq:E-field}
  E_\parallel\sim 4\pi\rho\lambda_p \sim \frac{4\pi j}{\omega_p},
\end{equation}
where
$\omega_p$ is the plasma frequency defined by
\begin{equation}
\label{eq:omega_p}
  \omega_p^2=\frac{4\pi e\rho}{m_e}, \qquad \rho=\frac{j}{c}.
\end{equation}
In other words, the surface charge near the anode and cathode is organized
so that particles extracted from the star are accelerated to $v\sim c$ over a
length comparable to the plasma skin depth.

For simplicity, consider a symmetric double layer where the positive and
negative charges have the same mass. In the 1D model, the electric field is
almost constant between the two surface charges of the double layer, giving a voltage drop,
\begin{equation}
  \label{eq:1}
  \frac{e\Phi_{e}}{m_{e}c^2} = \frac{4\pi j e L}{\omega_pm_{e}c^{2}} = \frac{\omega_p}{c}L = \frac{L}{\lambda_p},
\end{equation}
where $L$ is the size of the layer (the distance between the footpoints).
Using $j\sim \psi B/L$, one finds for the typical parameters of a magnetar,
\begin{equation}
  \label{eq:1D}
   \frac{L}{\lambda_p} \sim 10^{8}\; L_6^{3/2}\psi^{-1/2} B_{15}^{-1/2},
\end{equation}
which
implies a huge voltage $\Phi_e$.

The estimate in \Eq~(\ref{eq:1}) is not
valid, however, for a realistic
magnetosphere, which is not one-dimensional.
The current flows along the curved magnetic field lines and their dipolar geometry
significantly changes the distribution of
the net voltage
sustained between the two footpoints.

The corrected voltage may be estimated as follows.
Since $\lambda_p$ is small compared with the thickness of the j-bundle,
the surface charge remains thin and its structure is not changed from the 1D model.
The self-consistent electric field extracting charges from the footpoint is still described by
\Eq~(\ref{eq:E-field}). However, with increasing altitude the electric field
must be reduced on a scale comparable to the horizontal size of surface charge
$W$ (thickness of the j-bundle). The resulting potential drop saturates at
$\Phi_e\sim E_\parallel W$, which gives
\begin{equation}
  \label{eq:voltage}
  \gDL=\frac{e\Phi_{e}}{m_{e}c^2}
  \sim \frac{W}{\lambda_p}.
\end{equation}
It is smaller than the 1D estimate by the factor of $W/L$. For instance a
j-bundle of thickness $W\sim 0.1\Rs$ at the stellar surface and length $L\sim 10
\Rs$ would sustain a voltage $\sim 10^{-2}$ smaller than predicted by the 1D
model. This is still a huge voltage and particles that tap the full potential
drop will be able to induce pair discharge, making the double layer model
inconsistent.

One should also note that $E_\parallel=\bE\cdot\bB/B$, and hence the voltage,
\begin{equation}
\label{eq:phi_e}
  \Phi_{e} = \int_p^{q} E_\parallel\,d\ell,
\end{equation}
have a pure inductive origin. One should think of
$E_\parallel$ as $c^{-1}\partial A_\parallel/\partial t$, the result of the slow decay
of the ultrastrong twisted magnetic field (BT07).
$e\Phi_e$ measures the energy gain of charge $e$ completing the electric circuit,
and this released energy is extracted from the magnetic twist energy.
A potential electric field would be unable to support any significant voltage
between the footpoints, as they are connected by an excellent conductor --- the crust.

The induction electric field $\bE$ still satisfies the Gauss law $\nabla\cdot\bE=4\pi \rho$;
as long as the untwisting process occurs much slower than the light crossing of
the system, one can think of the dissipation as a quasi-steady process. The inductive
double layer is similar to a normal electrostatic double layer except that
the integral of $\bE$ along the full closed circuit (including the part
closing through the crust, where $\bE=0$) does not vanish and instead equals $\Phi_e$.
There is no external emf applied to the circuit below the stellar surface; the only emf
sustaining the current is the induction emf due to the twist decay in the magnetosphere
itself.

\subsection{Voltage with pair discharge}
\label{sec:v-with-pair}

The mechanism of secondary $e^\pm$ creation by relativistic particles in the
magnetar magnetosphere  involves an intermediate step of gamma-ray production.
It occurs through resonant Compton scattering of photons flowing from the star
by particles accelerated in the magnetosphere. A target photon with energy
$E_t\sim 1$~keV can be resonantly scattered by an electron with Lorentz factor
$\gamma$ if the photon energy measured in the electron rest frame matches $\hbar\omega_B$,
where $\omega_B=eB/m_ec$.\footnote{This simple resonance condition remains
   valid in ultrastrong fields $B\gg B_Q$ when one takes into account the electron recoil
   in scattering and the fact
   that
   the target photon is propagating almost parallel to $\bB$
   when viewed in the electron rest frame, because of the relativistic aberration effect (BT07).}
The resonance condition reads
\begin{equation}
  \label{eq:9}
  \gamma(1 - \beta\cos\theta_X)E_t = \hbar\omega_B,
\end{equation}
where $\theta_X$ is the angle of the target X-ray with respect to local magnetic
field line (the electron moves along the field line). The energy $\hbar\omega_B$
equals $m_ec^2$ for the characteristic magnetic field
$B_Q=m_e^2c^3/e\hbar\approx 4.4\times 10^{13}$~G and scales linearly with $B$.

Magnetars supply plenty of keV photons, and the electron Lorentz factor required
for resonant scattering at $B\sim B_Q$ is moderate, $\gamma \sim 10^{3}$. It is
far below the electron Lorentz factors that would be reached in the double layer
discussed in the previous section.

After the scattering, the photon energy is boosted by a factor comparable to
$\gamma^2$, putting the originally keV photon into the GeV range, $E_\gamma\sim
1$~GeV. Such energetic gamma-rays can easily convert to $e^{\pm}$ pairs in the
strong magnetic field, as soon as the gamma-ray pitch angle with respect to the
magnetic field, $\theta_\gamma$, is large enough to satisfy the threshold
condition,
\begin{equation}
  \label{eq:threshold}
  E_\gamma\sin\theta_\gamma > 2m_{e}c^2.
\end{equation}
In the region near the star where $B>10^{13}$~G the conversion occurs practically
immediately following resonance scattering \citep{2013ApJ...762...13B}.

The efficiency of pair creation implies a quick development of electric discharge
until the number of created particles becomes sufficient to screen the accelerating
electric field. The process develops in a runaway (exponential) manner and hence
the accelerating voltage is unlikely to grow beyond a characteristic value that
makes particles capable of resonant scattering. This condition defines a ``threshold''
for discharge, which corresponds to a characteristic electron Lorentz factor $\gthr$.

\subsection{Characteristic timescales and energy scales}

The shortest timescale of interest is the plasma scale $\omega_p^{-1}$. It
describes the growth rate of the local accelerating electric field in response
to charge starvation (BT07). It also determines the thickness of the surface
charge $c/\omega_p$ in the double-layer configuration.

The characteristic dynamic timescale of the electric circuit is the light crossing
time or the Alfv\'en crossing time of the system,
\begin{equation}
   \tA= \frac{L}{c}\sim 0.3\,L_7 {\rm~ms},
\end{equation}
where $L$ is the length of the magnetospheric field line. The group speed of Alfv\'en
waves is always directed along the magnetic field lines and its value is close to $c$
in the magnetically dominated corona.

The longest timescale in the problem is the lifetime of the magnetic twist.
The finite voltage sustaining the magnetospheric current implies a finite ohmic
dissipation rate, so the magnetic twist energy $E_\mathrm{twist}$ must dissipate with time,
\begin{equation}
  \label{eq:11}
  \frac{d\Etw}{dt}
  \approx
  \frac{d}{dt}\int\frac{B_{\phi}^2}{8\pi} \, dV \sim I\Phi_{e},
\end{equation}
where $I$ is electric current flowing through the magnetosphere.
The voltage $\Phi_e$ controls the timescale of this evolution,
\begin{equation}
  \tev\sim \frac{\Etw}{I\Phi_{e}}.
\end{equation}
Using the characteristic $I\simlt \psi (c/4\pi)B\Rs$ and $\gthr\sim 10^3$ one
can estimate that $\tev$ is comparable to one year. This theoretical timescale for
untwisting is comparable to the observed decay timescale in transient
magnetars following an outburst of activity.

Because of the vast separation of timescales, $\tev\gg \tA$, the ohmic dissipation
of the magnetospheric twist can be viewed as a quasi-steady process
slowly draining the twist energy. Unsteadiness of the discharge may lead to
strong variability in the electric circuit, however
it occurs
on very short timescales, which
would be hard to resolve observationally.

The characteristic scales for energy (or electron Lorentz factor $\gamma$) also have
an important hierarchy. The highest energy corresponds to $\gDL$, which would only
be achieved in the absence of pair creation.
It
is given by \Eq~(\ref{eq:voltage})
and can
exceed $10^6$. The next characteristic $\gamma$ is determined by the threshold for
$e^\pm$ discharge $\gthr$, which is comparable to $10^3$. Both $\gDL$ and $\gthr$
are much greater than unity.

\subsection{Mechanism of untwisting}
\label{sec:twist-theory}

An integral form of the Faraday's induction law $\partial \bB/\partial t=-c\nabla\times\bE$
leads to a simple equation describing
resistive evolution of the
axisymmetric twist
\citep{2011ASSP...21..299B},
\begin{equation}
  \label{eq:twist-ev}
  \dot{\psi} = 2\pi c \frac{\partial\Phi_{e}}{\partial f}.
\end{equation}
Here $f(r,\theta)$ is the poloidal magnetic flux function (constant along a magnetic flux
surface), which serves to label the magnetic field lines. For any given point $(r,\theta)$,
$f$ is defined as the magnetic flux through the circle about the axis of symmetry
passing through the point; $f = 0$ on the axis of symmetry.
In particular, for a dipole poloidal field with a dipole moment $\mu$ the flux
function is given by
\begin{equation}
  f = \frac{2\pi\mu \sin^2\theta}{r}, \qquad 0\leq f \leq \fmax=\frac{2\pi\mu}{\Rs}.
\end{equation}
Note that $\sin^2\theta/r=\mathrm{const}$ along a dipole field line. It is convenient to use the
dimensionless flux function
\begin{equation}
\label{eq:u}
  u\equiv\frac{f}{\fmax}=\sin^2\theta_\star,
\end{equation}
 where $\theta_\star$ is the polar angle of the magnetic field line footprint on the stellar
surface.

Equation \eqref{eq:twist-ev} shows that the twist must decrease where
$\partial\Phi_{e}/\partial f < 0$ and increase where $\partial\Phi_{e}/\partial
f > 0$. The fact that $\Phi_e(\fmax)=0$ (the field line $\fmax$ is confined to
the star, which we approximate as an ideal conductor) implies
$\partial\Phi_e/\partial f<0$ at some $f<\fmax$. This region with large $f$,
comparable to $\fmax$, corresponds to the inner magnetosphere near the equator,
with short field lines. B09 showed that this fact leads to immediate formation
of a ``cavity'' with $j=0$ in the equatorial region near the star, and the
cavity expands on the timescale $\tev$, erasing the magnetospheric currents. The
currents are ``sucked'' into the star, so that they close inside the conductor.

From the untwisting equation it is evident that the profile of $\Phi_e(f)$ plays the
key role for the twist evolution. Voltage regulated by pair discharge is expected to
satisfy the condition $e\Phi_e\sim \gthr m_ec^2$.
Its variation with $f$ over a region $\Delta f=\fmax \Delta u$ gives the characteristic
twist evolution timescale,
\begin{equation}
  \label{eq:tev}
  \tev=\frac{\psi}{\dot{\psi}} \sim
  \frac{\mu}{c\,\Phi_\mathrm{thr}\Rs}\,\psi\,\Delta u.
\end{equation}
The dimensionless quantities $\Delta u$ and $\psi$ are comparable to unity,
and the characteristic timescale is set by the ratio $\mu /\Phi_\mathrm{thr}$.
Note however that $\tev$ can strongly differ for different magnetic field lines.
In particular, if there is a region with a flat dependence of $\Phi_e(f)$,
$\partial\Phi_e/\partial f=0$,
then the local $\tev=\infty$ and the twist angle $\psi$ is ``frozen'', waiting for the cavity
expansion to reach the region (B09).

Another interesting implication of \Eq~(\ref{eq:twist-ev}) is that on some field
lines the twist may {\it grow} as the magnetosphere untwists. In particular, a
decrease of $\Phi_e$ toward the magnetic axis, $\partial\Phi_e/\partial f>0$,
leads to $\dot{\psi}>0$. This effect will be observed in the simulations below.
Together with the cavity expansion, this means that the twist relocates toward
the axis with a decreasing energy $\Etw$ but possibly
with
increasing amplitude $\psi$ in some regions before being completely dissipated.

\section{Setup of the simulation}
\label{sec:setup}

\subsection{Implanting the twist}
\label{sec:implant}

Our simulation starts with a pure dipole magnetosphere, with a magnetic moment
${\mathbf \mu}$ and no magnetic twist, $B_\phi=0$. The twist is gradually
implanted by shearing the stellar surface with a latitude-dependent angular
velocity $\boldsymbol{\omega}(\theta)\parallel\boldsymbol{\mu}$. The profile of
$\omega(\theta)$ determines the profile of the implanted twist; we choose a
profile similar to previous magnetohydrodynamic (MHD) and force-free
electrodynamic (FFE) simulations of twisted magnetospheres
(\citealt{1994ApJ...430..898M}; PBH13),
\begin{equation}
  \label{eq:3}
  \omega(\theta,t) = \omega_0(t)\frac{\Theta}{\sin\theta}\exp \left[ (1 - \Theta^4)/4 \right],
\end{equation}
where $\Theta = (\theta - \pi/2)/\Delta\theta_m$ and $\Delta \theta_m = \pi/4$
is a measure of the width of the sheared region. This profile gives a smooth
twist that is centered at $\theta=\pi/4$ and decreases to zero at the equator.
The prefactor $\omega_0(t)$ describes the rate of implanting the twist. It is
smoothly increased from zero at $t=0$ to a chosen maximum value, kept at this
value for some time, and then smoothly switched off back to zero.

As long as the duration $\ttw$ of the surface shear $\omega\neq 0$ is shorter than
the resistive timescale of the magnetosphere, $\ttw\ll\tev$, ohmic dissipation
may be neglected during time $\ttw$. Then the implanted twist profile is given by
\begin{equation}
  \psi(\theta)=\int_0^{\ttw}\omega(\theta,t)\,dt.
\end{equation}
We choose $\ttw=40 \Rs/c$. Then the shearing stage is sufficiently short
compared with the total duration of our simulation $t_\mathrm{sim} = 350\Rs/c$
but longer than or comparable to the Alfv\'en crossing time $\tA$ of the sheared
region, so that twist implanting is a relatively gentle process. The maximum
shear angle (near $\theta=\pi/4$) is $\psi_{\max}\approx 1.6$~radian in the
simulations presented below.

After the twist implantation is finished, $\omega$ is kept at zero and the
boundary condition at the stellar surface becomes simply a perfect static
conductor. Magnetars are slow rotators, and their light cylinders $R_{\rm
  LC}\simgt 10^4\Rs$ are well beyond the twisted, dissipative region. The slow
spinning of the star is neglected in the present paper, which corresponds to
$R_{\rm LC}=\infty$.

The implanted twist $\psi\sim 1$ is moderate and expected to result
in moderate inflation of the poloidal magnetic field lines. The main effect of surface
shearing is creating a strong $B_\phi$ in the magnetosphere.
Analytical arguments
\citep[e.g.][]{2002ApJ...574.1011U} and FFE simulations (PBH13) show that a
stronger $\psi \simgt 3$ will result in a global instability of the
magnetosphere, which we do not intend to study in this paper and defer to future
work.

\subsection{Surface atmospheric layer}
\label{sec:atm}

We start the simulation with a complete vacuum around the star and create
a dense neutral atmospheric layer at the stellar surface by injecting warm electron-ion
plasma at $\Rs$.
The atmosphere scale-height $h$ is determined by the particle injection temperature and
gravity of the star. We choose a Maxwellian injection velocity with the mean value
$v_0\approx 0.1c$ and the gravitational acceleration $g = g_{0}/r^2$ with
$g_0 = 0.5\Rs c^2$. This gives the hydrostatic scale-height
\begin{equation}
  h\approx\frac{v_0^2}{2g_0}\approx 0.01\Rs.
\end{equation}
This is a much thicker atmospheric layer than the magnetar would have at a
surface temperature $kT\simlt 1$~keV. However, it is sufficiently thin and still
resolved by our numerical grid (see below).
The characteristic time it takes to form the atmosphere is short,
$t_{\rm atm}\sim h/v_0= 0.1 \Rs/c$. Throughout the simulation particles are continually
injected and absorbed by the star, sustaining a steady atmosphere at $t\gg t_{\rm atm}$.

The injection rate is chosen high enough to ensure a high density at the base of
the atmosphere,
\begin{equation}
  n_{\rm atm}\gg \frac{j}{ev_0}.
\end{equation}
The density is exponentially reduced with altitude on the scale $h$, and steeply
drops to a low value below $j/ec$. Therefore, in the absence of $E_\parallel$ the
hydrostatic plasma is not capable of conducting the electric current $j$ required in
the twisted magnetosphere.

Where the atmospheric density $n(r)$ falls below $j/ec$, electric field
$E_\parallel$ is expected to develop in response to charge starvation and lift
particles from the atmosphere. The thin and dense atmospheric layer merely makes
plasma available, with no special injection assumptions at the stellar surface.
The numerical experiment must show how the system responds to the surface shear
described in Section~\ref{sec:implant} and whether the induced $E_\parallel$
will self-organize to conduct the magnetospheric currents that allow the twist
to be implanted.

\subsection{Creation of $e^\pm$ pairs}
\label{sec:pairs}

If $E_\parallel$ accelerates the lifted electrons to high Lorentz factors
$\gamma>\gthr$, pair creation will be ignited. In this paper, we use the
simplest implementation of this process: we choose a fixed value for $\gthr$ and
let a new $e^\pm$ pair be instantaneously created every time an electron (or
positron) reaches $\gthr$. This may be a reasonable approximation for the
$e^\pm$ discharge near the star where $B\gg 10^{13}\,\mathrm{G}$
\citep{2013ApJ...762...13B}. However, it becomes poor at larger distances where
the magnetic field is weak and resonantly scattered photons have lower energies.

An additional simplification in our implementation is the
prescription for the energy of the created pair. We will assume that the pair takes a
fixed energy $\Delta E$ from the primary particle, and shares it equally, i.e. the new
$e^+$ and $e^-$ each receives $\Delta E/2$ (including the rest mass).
Total energy and momentum parallel to $\bB$ is conserved in the pair creation process.

Thus, we do not track the propagation of any high-energy photons, which is
significantly simpler than the discharge model of CB14 developed for pulsars.
The simplified version appears adequate for the first axisymmetric PIC model of
magnetars. It should be sufficient to demonstrate some basic features of plasma
self-organization in response to shearing of the magnetospheric footpoints,
followed by ohmic dissipation of the twist. The results may be used as a
benchmark for future more advanced simulations. Future simulations will have
explicitly implemented
resonant scattering process
, so that $\Delta E$ will be the energy of
the resonantly scattered photon, which may convert to $e^\pm$ with a delay. Both
$\gthr$ and $\Delta E$ will vary with the local magnetic field, see
\citet{2013ApJ...762...13B}
for a detailed discussion.

\subsection{Rescaling of large numbers in the problem}
\label{sec:rescale}

Any PIC simulation must resolve the plasma skin depth $\lambda_p=c/\omega_p$,
which is a demanding condition on the computational grid, as $\lambda_p$ is a
microscopic scale and the ratio $\Rs/\lambda_p$ is huge (comparable to $10^8$ in
magnetars). Similar to the PIC simulations of rotation-powered pulsars, this
issue is resolved by rescaling the parameters of the problem so that $\lambda_p$
remains much smaller than the stellar radius, $\lambda_p\sim 10^{-2}\Rs$,
but becoming
sufficiently large to be well resolved. This rescaling has two
main implications:
\\
(1) Similar to the pulsar problem, the increased $\lambda_p$ implies a reduction
of the energy scales (cf. CB14). In particular, the maximum voltage that can be
induced in a magnetar magnetosphere is given by $\gDL$ (\Eq~\ref{eq:voltage}),
which now becomes moderate, $\gDL\sim 10^2$. To respect the hierarchy of the
energy scales $1\ll\gthr\ll\gDL$, a good choice for the discharge threshold in
the numercial experiment is $\gthr\sim 10$. Secondary pairs receive the energy
$\Delta E$, which must be a fraction of $\gthr m_ec^2$. We will fix $\Delta
E=3.5 m_ec^2$ for all simulations presented below.
\\
(2) The rescaling of $\lambda_p$ changes the lifetime of the implanted twist, as
seen from the following estimate. The value of $\lambda_p=c/\omega_p$ is related
to the electric current density $j$ by \Eq~(\ref{eq:omega_p}), and the
characteristic value of $j$ scales with the magnetic dipole moment of the star
$\mu$: $j\sim \psi\, (c/4\pi)(\mu/\Rs^4)$. This gives, \beq
\label{eq:mu}
  \left(\frac{\lambda_p}{\Rs}\right)^2\sim \frac{m_ec^2\Rs^2}{e\mu\psi}.
\eeq
Combining this relation with \Eq~(\ref{eq:tev}) for the resistive evolution timescale, one obtains
\beq
\label{eq:tev1}
   \tev\sim\gthr^{-1}\left(\frac{\Rs}{\lambda_p}\right)^2\frac{\Rs}{c}.
\eeq
One can see that the rescaling of $\lambda_p$ to $\sim 10^{-2}\Rs$ reduces the
resistive timescale to $\tev\sim 10^3(\Rs/c)$ when $\gthr\sim 10$. This is fortunate,
as the untwisting evolution can now be observed during a reasonably long simulation.
With the realistic $\lambda_p/\Rs\sim 10^{-8}$ and $\gthr\sim 10^3$
one would have $\tev\sim 10^{13}\Rs/c$.

Another large number that should
be
rescaled in the simulation is the ion-to-electron
mass ratio $m_i/m_e\approx 2\times 10^3$. We use $m_i/m_e=10$.
This rescaling is useful for two reasons:
(1) The characteristic ion plasma frequency $\omega_{p,i}=(4\pi n_i e^2/m_i)^{1/2}$
is not very much smaller than $\omega_p$, so that $\omega_{p,i}<r/c$ is well
satsified, and (2) $m_ic^2$ becomes comparable to
$\gthr m_ec^2$. The latter coincidence is also expected for the real magnetar discharge.

It is also useful to evaluate the surface magnetic field $\Bs\sim\mu/\Rs^3$, which can
be expressed from \Eq~(\ref{eq:mu}), and then estimate the characteristic gyro-frequency,
\beq
   \omega_B=\frac{e\Bs}{m_ec}\sim \frac{c}{\Rs}\left(\frac{\Rs}{\lambda_p}\right)^2,
\eeq
where $\lambda_p$ corresponds to the current density supporting a twist $\psi\sim 1$.
One can see that the particles are very strongly magnetized, $\omega_B\sim 10^4 c/\Rs$,
and hence expected to move along the magnetic field lines, similar to real magnetars.
The characteristic gyro-frequency is also related to another important parameter of
the magnetosphere --- the ratio of magnetic and plasma energy densities,
\beq
   q=\frac{B^2}{4\pi\gamma n m_ec^2}=\frac{\omega_B^2}{\gamma\omega_p^2}.
\eeq
For real parameters of magnetars this ratio is $q\sim 10^{17}$.
The characteristic parameters chosen in our simulations give $q\sim 10^3$.
This is still very much above unity, so the magnetosphere is nearly force-free as
it should be.

The parameter $q$ also determines the Lorentz factor of Alfv\'en waves,
$\gamma_{\rm A}\approx q^{1/2}$. For a real magnetar, this gives
$\gamma_{\rm A}\gg \gamma\sim\gthr$.
This condition is satisfied in our rescaled numerical experiment
 as long as $\gthr\ll 30$.

\subsection{Evolving the fields and the plasma: Aperture}
\label{sec:aperture}

The particle-in-cell (PIC) method provides an efficient technique to simulate plasma
from first principles. The electromagnetic fields are evolved on a grid according to
Maxwell equations with the source (electric current and charge density) provided by
the plasma that is self-consistently evolved in the electromagnetic field. The plasma is
represented directly as a large number of individual particles. The simulation follows
the motion of each particle by calculating the applied forces.
The motion of the plasma particles creates electric current which is interpolated onto the
grid and then used as the source term in the Maxwell equations to update
the electromagnetic field. The method well describes the plasma behavior at the
microscopic kinetic level as long as the plasma skin depth is well resolved by the grid
and the number of particles per grid cell is much larger than one.

Our simulations are performed using the PIC code Aperture.\footnote{Aperture is
  a recursive acronym:
  Aperture is a code for Particles, Electromagnetic
  fields, and Radiative Transfer at Ultra-Relativistic Energies.} The code was
originally developed for the PIC simulations of rotationally powered pulsars
(CB14). The code can follow pair creation with or without explicit tracking of
high-energy photons. In the present work we use the simplified implementation of
pair creation (Section~\ref{sec:pairs}) and do not use the radiative transfer
module. The code is fully relativistic and designed to work on curvilinear
grids. This is particularly important for problems with natural spherical
geometry, such as the plasma dynamics around a spherical star in a region
extending far beyond the stellar radius.

The simulations presented below are done in 2.5D, which means that our grid is
2D (in the poloidal plane) but all vector quantities are fully 3D, and we solve
the full Maxwell equations assuming axisymmetry. Particles in the simulation may
be thought of as rings with poloidal and toroidal velocity components. We use a
spherical $r,\theta$ grid with logarithmic spacing in $r$ and uniform spacing in
$\theta$. For all of the simulations shown in this paper, the grid size is
$384\times 384$ and the timestep $\Delta t = 10^{-3}\Rs/c$.

The outer boundary of the simulation box is set at $\rout = 30\Rs$ and employs
a damping condition that lets outgoing electromagnetic waves and particles escape
the box, preventing reflection.
We did not detect any appreciable reflection of waves from the outer boundary.
Note also that most of the active (current carrying) field lines are
closed well inside the box and do not cross the outer boundary.

The shear motion of the stellar surface during the twist implantation stage
$t<\ttw=40\Rs/c$ is equivalent to imposing a tangential electric field at the
boundary. The field corresponding to the surface motion with velocity $\bv$ in
the lab frame is given by $\bE=-\bv\times\bB/c$. It corresponds to zero electric
field in the comoving frame of the stellar crust, which is assumed to be an
ideal conductor. This gives the following boundary condition at $r=\Rs$,
\begin{equation}
  \label{eq:4}
  \mathbf{E}(t,\theta) =
  -\frac{( \boldsymbol{\omega}(t,\theta)\times \mathbf{r})\times \mathbf{B}}{c}.
\end{equation}
The initial state is a dipole field and the normal component of the magnetic field
at the surface remains unchanged during the simulation.

\subsection{Units}
\label{sec:units}

A set of natural units can be defined as follows.
All lengths are measured in units the stellar radius $\Rs$ and time
is measured in $\Rs/c$. The corresponding velocity unit is the speed of light $c$.
We define the dimensionless electromagnetic field and current density as
\begin{equation}
  \label{eq:dimensionless}
  \tilde{E} = \frac{e\Rs E}{m_ec^2},\quad \tilde{B}
    = \frac{e\Rs B}{m_ec^2},\quad \tilde{\jmath} = \frac{4\pi e\Rs^2j}{m_ec^3}.
\end{equation}
Hereafter we will use tilde to denote dimensionless quantities,
e.g. $\tilde r=r/\Rs$, $\tilde{t}=ct/\Rs$ etc.

\section{Results}
\label{sec:result}

In all simulations presented below the magnetic field strength at the pole of
the star is $\tilde{B}_\mathrm{pole} = 4\times 10^{4}$. It corresponds to
$\tilde{\omega}_B=4\times 10^4$. We focus on the simulation with $\gthr=10$, as it
gives the best re-scaled model of real magnetars (Section~\ref{sec:rescale}). Simulations with
different $\gthr$ are only discussed in Section~\ref{sec:voltage}.

\subsection{Initial relaxation}

During the initial stage of the simulation $\tilde{t}<\tilde{t}_{\rm shear}=40$
the dipole magnetosphere is twisted by the surface shearing motion described in
Section~\ref{sec:implant}. The surface motion induces a parallel electric field
$E_{\parallel}$, which lifts charges from the atmospheric layer into the
magnetosphere and accelerates them. The electron Lorentz factors quickly reach
$\gthr$ and $e^{\pm}$ discharge is triggered within a single Alfv\'en time of
the twisted field line bundle.

The $e^\pm$ plasma created by the discharge screens $E_\parallel$, and the voltage
along the current loop temporarily drops, shutting down the discharge.
As  the created pairs are lost to the star on the light-crossing time,
a charge-starved region with significant $E_\parallel$ develops again.
This first happens near the equatorial plane.
As a result, an equatorial gap with strong $E_\parallel$ emerges and begins to
accelerate particles, sustaining the pair creation process. The gap structure and how
the $e^\pm$ discharge is sustained will be described in more detail in Section~\ref{sec:gap}.

It is clear from the simulation that a magnetospheric source of pair plasma is
established in the twisted magnetosphere on a timescale not much longer than the
light crossing time, before the surface shearing ends at $\ttw$. Pair creation
becomes the dominant source of plasma; the extraction of particles from the
atmospheric layer is only important at the initial stage igniting the $e^\pm$
discharge. After the pair discharge is activated, only a small fraction of the
magnetospheric current is carried by the particles lifted from the surface. In
particular, we observed that less than 1\% of the current is carried by the
ions.

We also observed that the twist implantation at $t<\ttw$ is accompanied by excitation
of Alfv\'en waves, which bounce back and forth along the magnetospheric field
lines.\footnote{Alfv\'en waves
    are reflected from the rigid sphere and trapped in the magnetosphere. Our simulation
    neglects the fact that the crustal material has a finite strength, which can lead to
    plastic damping of Alfv\'en waves in the crust
    \citep{2015ApJ...815...25L}.
  }
Similar waves were observed in FFE simulations (PBH13).
The waves are damped in the magnetosphere at later times, and the initial relaxation
period is followed by the gradual evolution on a much longer timescale $\tilde{t}_\mathrm{ohm}\gg 100$.

After the surface shearing stopped at $\ttw$, the electric discharge
persisted for the rest of the simulation. It continually supported the electric current in
the slowly untwisting magnetosphere, and the created particles continually bombarded the star.
The duration of the simulation $\tilde{t}_{\rm sim}=350$ was approximately 9 times longer than
$\ttw$ and comparable to the expected resistive timescale $\tev$ estimated in Section~\ref{sec:theory}.
The observed gradual evolution of the magnetospheric twist and currents on the
timescale $\sim\tev$ will be described in Section~\ref{sec:untwisting}.

\subsection{The equatorial gap}
\label{sec:gap}

\begin{figure}[t]
  \centering
  \includegraphics[width=0.5\textwidth]{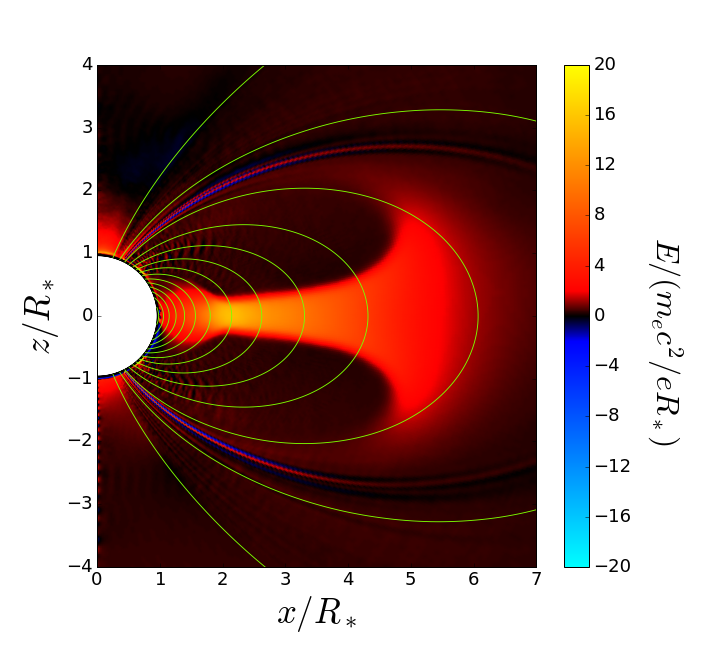}
  \caption{Electric gap in the twisted magnetosphere. Magnetic field lines
   are shown by the green curves (poloidal cross section), and color shows the
    parallel electric field, defined as $E_\parallel=\mathbf{E}\cdot \mathbf{B}/B$,
    in our standard units defined in Section~\ref{sec:units}.
    The plot shows the average of a series of snapshots centered around
    $\tilde{t} = 200$. The gap voltage is self-regulated to the discharge threshold
    $\gthr$;  $\gthr=10$ in the simulation.}
  \label{fig:EdotB}
\end{figure}

A key aspect of the discharge self-organization is how and where
particles are accelerated. The simulation clearly shows the formation of a quasi-steady
``gap'' with a strong $E_\parallel$ concentrated around the equatorial plane
(Figure~\ref{fig:EdotB}). The gap thickness $\lgap$ is smaller than radius,
and its voltage is near the threshold for $e^\pm$ discharge,
\begin{equation}
  \Phigap\approx\lgap\,\Egap, \qquad e\Phigap\approx\gthr m_ec^2.
\end{equation}
Particles are accelerated in the gap and most of the pair
creation events happen around this region.

As seen in Figure~\ref{fig:EdotB}, the gap has a rather sharp boundary;
$E_\parallel$ is screened outside it by the created $e^\pm$ plasma.
The drop of $E_\parallel$ across the two boundaries of the gap is sustained by the layers
of positive and negative charge ($\pm\Sigma$ above and below the equatorial plane,
respectively), according to Gauss law $\nabla\cdot\bE=4\pi\rho$.
The charged layers are self-consistently sustained by the difference in velocities of
positive and negative charges passing through them in the self-organized $E_\parallel$.

\begin{figure}[t]
  \centering
  \includegraphics[width=0.5\textwidth]{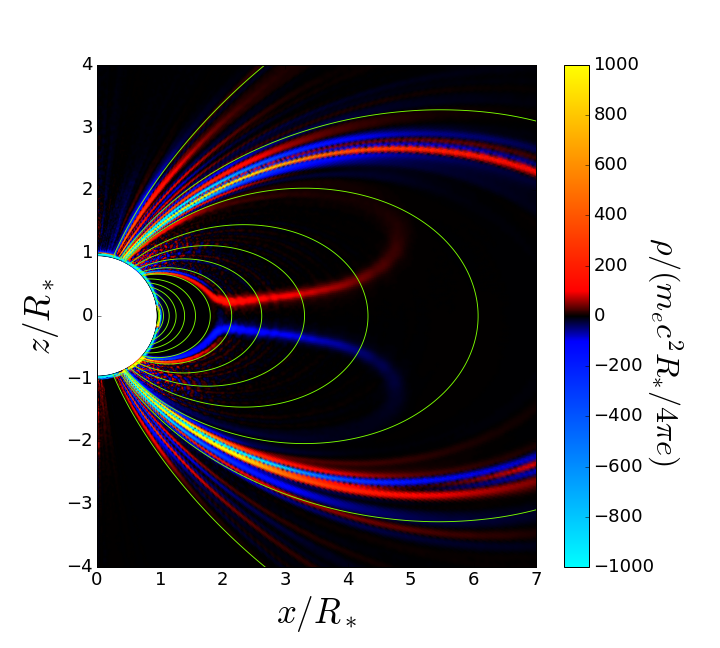}
  \caption{Charge density in the magnetosphere, averaged in the same way as
      in Figure \ref{fig:EdotB}. Note the thin charged
  layers bounding the equatorial gap
  across the magnetic field lines. The layers extend into the inner magnetosphere
  along the inner boundary of the j-bundle. The charged structure observed on the field
  lines extending to $\tilde{r}\sim 9$ approximately corresponds to the outer boundary of
  the j-bundle (see Figure~\ref{fig:current}).}

  \label{fig:rho}
\end{figure}

In essence, the gap is a double layer. It has been compressed toward
the equatorial plane to a minimum thickness $\lgap$ that is still capable of sustaining
particle acceleration to $\gthr$.
Similar to the double layer described in Section~\ref{sec:v-no-pair}, the charge layers sandwiching
the gap have the thickness comparable to the local plasma skin depth $\lambda_p$
(evaluated for charge density $\sim j/c$) (Figure \ref{fig:rho}).
The electric field in the gap is $\Egap\sim 4\pi (j/c)\lambda_p$ and its voltage is
\begin{equation}
  e\Phigap\sim \frac{\lgap}{\lambda_p}\,m_ec^2.
\end{equation}
The self-regulation of the gap voltage to $\Phigap\approx \Phi_{\rm thr}$
controls the gap thickness $\lgap\sim\gthr\lambda_p$.

Unlike normal double layers, particles accelerated in the gap are not brought
from outside; instead, the gap feeds itself with particles. The accelerated
particles create secondary $e^\pm$ of lower energies near the gap exit, and some
of the secondary particles are reversed by $E_\parallel$ and accelerated toward
the opposite boundary of the gap, where they create new pairs, etc.

\begin{figure}[t]
  \centering
  \includegraphics[width=0.5\textwidth]{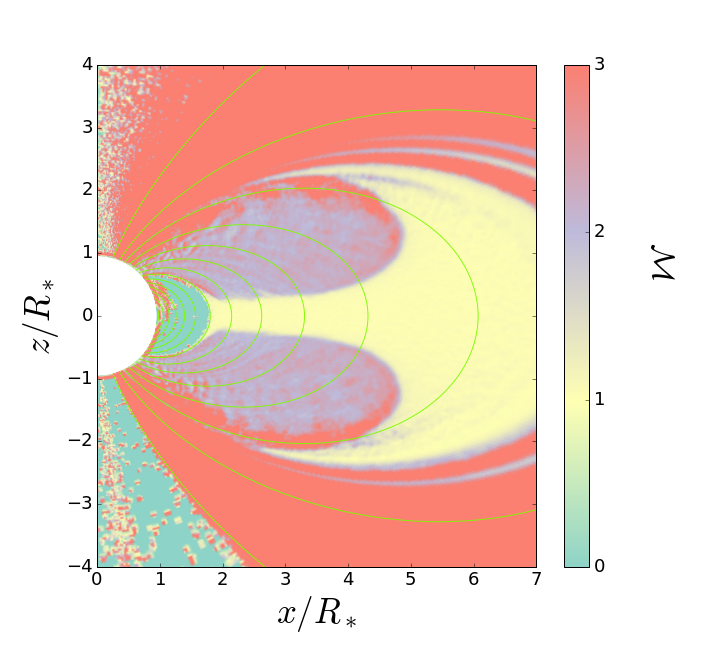}
  \caption{Pair multiplicity $\mathcal{M} = (\rho_+ - \rho_{-})/j$.}
  \label{fig:multiplicity}
\end{figure}

The multiplicity of the pair plasma is defined by $\mathcal{M} = (\rho_+-\rho_{-})/j$,
where $\rho_+$ and $\rho_-$ are the charge densities of the positrons and electrons,
respectively. One can see in Figure~\ref{fig:multiplicity} that $\mathcal{M}$
in the gap is close to 1, i.e. the gap contains the minimum amount of plasma needed to
conduct the electric current.
This is consistent
with no screening in the gap that allows the strong $E_\parallel$ to be
sustained. Pair multiplicity in other parts of the j-bundle is
close to 2,
just sufficient to screen $E_\parallel$. Apparently, the discharge in the
simulation is self-organized to carry the current with the minimum voltage
$\Phi_e\approx\Phigap\approx\Phi_{\rm thr}$ and the minimum rate of pair
creation.

\begin{figure*}[t]
  \centering
  \includegraphics[width=0.9\textwidth]{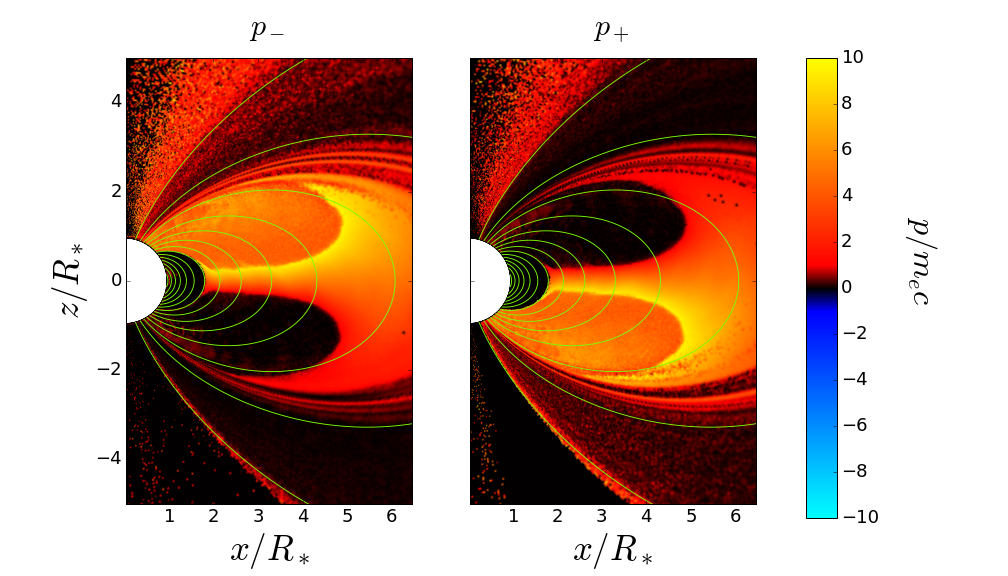}
  \caption{Average hydrodynamic momentum of electrons (left) and positrons (right).}
  \label{fig:momenta}
\end{figure*}

Figure~\ref{fig:momenta} shows the average hydrodynamic momenta of
electrons and positrons. It is apparent that both species are accelerated across the
equatorial gap to the threshold Lorentz factor $\gthr=10$. The move with almost
speed of light in the opposite directions and make approximately equal contributions
to the current density, consistent with $\M\approx 1$.
Outside the gap,
$\M\approx 2$ together with the charge neutrality condition $n_+\approx n_-$
implies that the current is carried by one species while the other creates the
neutralizing, nearly static, background. This is indeed observed in
Figure~\ref{fig:momenta}.

The gap voltage is not exactly steady and shows quasi-periodic ``breathing'' with time.
This must assist the gap in reversing some of the secondary particles so that
they can cross the gap and accelerate to $\gthr$, sustaining the pair creation cycle.
Most of the accelerated particles escape the gap and get absorbed by the star.

Since the magnetosphere was set up to be symmetric about the equatorial plane,
the fact that the current is strongly dominated by created pairs  implies
symmetric bombardment of the two footprints of the j-bundle. Thus, our simulation
shows two symmetric hot spots (or rather rings, due to the axial symmetry) in the
northern and southern hemispheres of the star.

As discussed in BT07 and Section~\ref{sec:v-no-pair}, the voltage $\Phi_e$ in
the magnetospheric circuit is purely inductive. The parallel electric field
$\bE= -c^{-1}\partial \mathbf{A}/\partial t$ is associated with the slow
dissipation of $B_{\phi}$ rather than an electrostatic potential.
Note also that the dissipation rate $\bE\cdot\bj=E_\parallel j$ is localized in the
gap while the untwisting of $B_\phi$ also occurs outside the gap. The re-distribution of the
dissipated $B_\phi$ along the j-bundle into the screened region with $E_\parallel\approx 0$
occurs through the Alfv\'en mode, which can propagate without dissipation.
The Alfv\'en timescale $\tA\sim r/c$ is much shorter than the untwisting timescale $\tev$,
and so the magnetosphere slowly evolves through the sequence of global twist equilibria
of a decreasing energy
$\Etw$,
even though the magnetic energy is converted to heat only
near the equator.

\subsection{Dependence on the threshold voltage}
\label{sec:voltage}

While the simulation with $\gthr=10$ is the most adequate re-scaled version of the
magnetar magnetosphere (Section~\ref{sec:rescale}), we also performed simulations with
$\gthr=20$, 100, and $\infty$ (no pair creation). All other parameters of the four
simulations were identical.

\begin{figure}[t]
  \centering
  \includegraphics[width=0.5\textwidth]{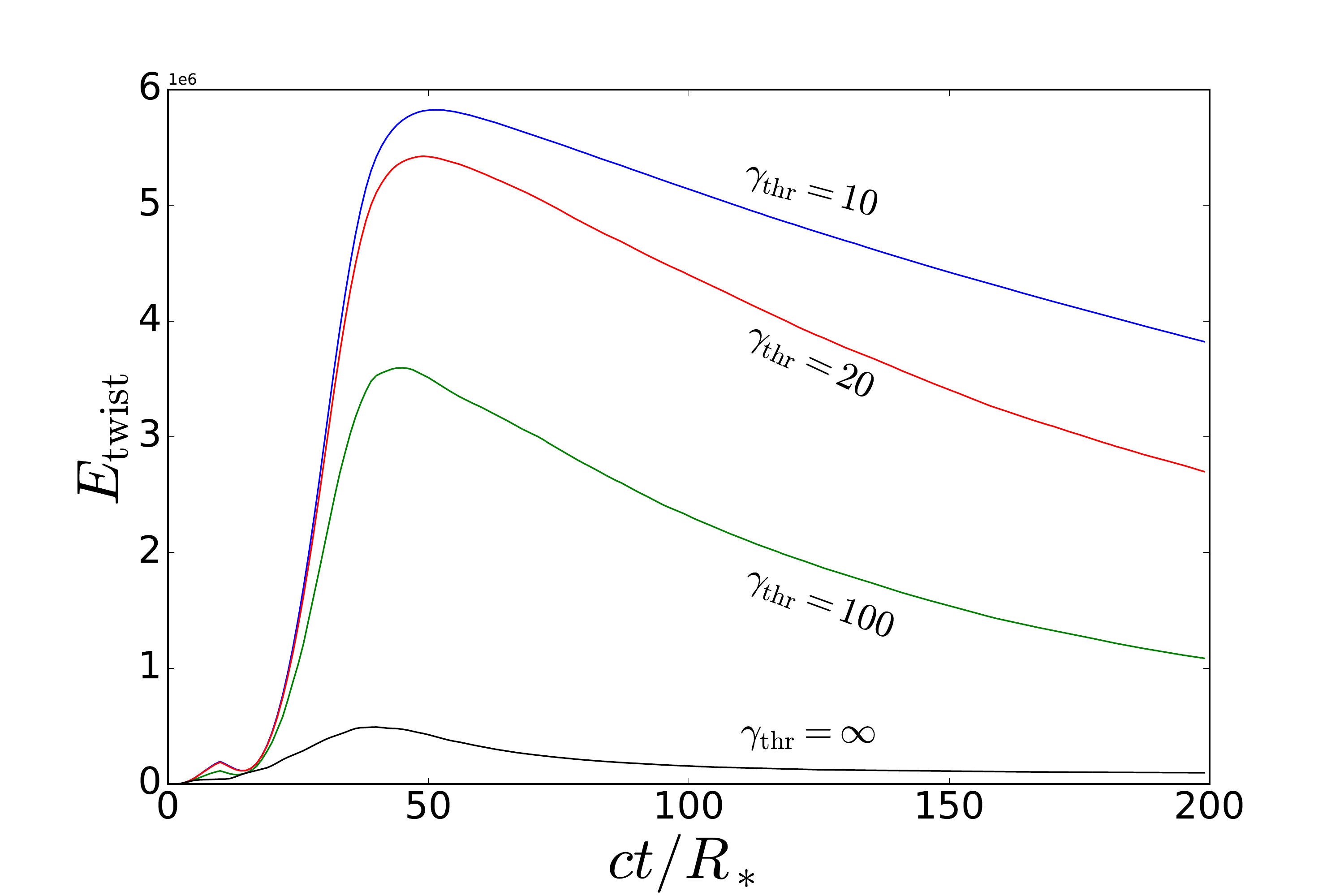}
  \caption{Evolution of the twist magnetic energy $\Etw$.
     Four simulations are shown with discharge thresholds
    $\gthr=10$, 20, 100, $\infty$. We use the exact expression for
    $\Etw = \int (B^2 - B_0^2)/8\pi\, dV$,
    where $B_0$ is the initial dipole field. It takes into account that besides $B_\phi^2/8\pi$
    part of the twist energy is stored in the inflated poloidal magnetic field, which
    becomes important when the twist amplitude $\psi$ exceeds unity.
  }
  \label{fig:b-energy}
\end{figure}

Figure \ref{fig:b-energy} shows the evolution of the twist energy $\Etw$
in the simulations with the four different values of $\gthr$. An obvious trend is observed:
a higher threshold voltage for discharge, $e\Phithr=\gthr m_ec^2$, leads to a higher
dissipation rate and a shorter lifetime of the magnetic twist. When $\gthr\gg 10$,
the dissipation becomes so strong that it affects the initial stage of the twist
implantation at $\tilde{t}<\tilde{t}_\mathrm{shear}=40$, so that a substantial part of the twist amplitude
(and the corresponding energy $\Etw$) is lost before it could be implanted.

The extreme model with $\gthr=\infty$ gives so strong dissipation that $\Etw$
does not reach even 10\% of its target value. It is instructive to compare this
simulation with the expected dissipation rate in the pair-free configuration
described in Section~\ref{sec:v-no-pair}. From equation \eqref{eq:voltage}, we
can estimate the voltage drop of the double layer as $\gDL=\tilde{\Phi}_{e} \sim
\sqrt{\tilde{\jmath}}\,\tilde{W}$. The initial width of the j-bundle near the
star is $\tilde{W}\sim 1$. The target current density reaches $\tilde{\jmath}
\sim 3\times 10^4$ if the twist is fully implanted. This estimate gives $\gDL$
comparable to $200$; the actual voltage in the simulation reaches somewhat
higher values. The high voltage develops early during the shearing stage and
results in strong dissipation, which does not allow $\tilde{j}$ to approach
$3\times 10^4$.

The simulation with $\gthr=100$ enables the pair discharge, which buffers the
voltage growth in the j-bundle and allows a stronger twist to be implanted.
The simulations with $\gamma_\mathrm{thr}=20$ and, in particular, $\gthr=10$,
allow almost full implantation of the target twist with small ohmic losses. The subsequent  slow resistive evolution is similar in the two models, as both have $\Phithr$ well below
the double-layer voltage and sustain a long-lived discharge activity in the j-bundle.
As expected, the untwisting timescale $\tev$ is reduced by a factor of $2$ as
$\gthr$ is increased from 10 to 20 (see \Eq~\ref{eq:tev}).

These results unambiguously demonstrate that the energy dissipation timescale is
controlled by the pair creation threshold, confirming the conclusion of BT07.
In real magnetars, we expect $\gthr\ll\gDL$ (Section~\ref{sec:theory}).
Therefore, the most relevant model is the one with low $\gthr=10$, which is still high
enough to accelerated particles to ultra-relativistic energies and produce relativistic
secondary $e^\pm$.

\subsection{Expanding cavity}
\label{sec:untwisting}

\begin{figure*}[t]
  \centering
  \includegraphics[width=0.8\textwidth]{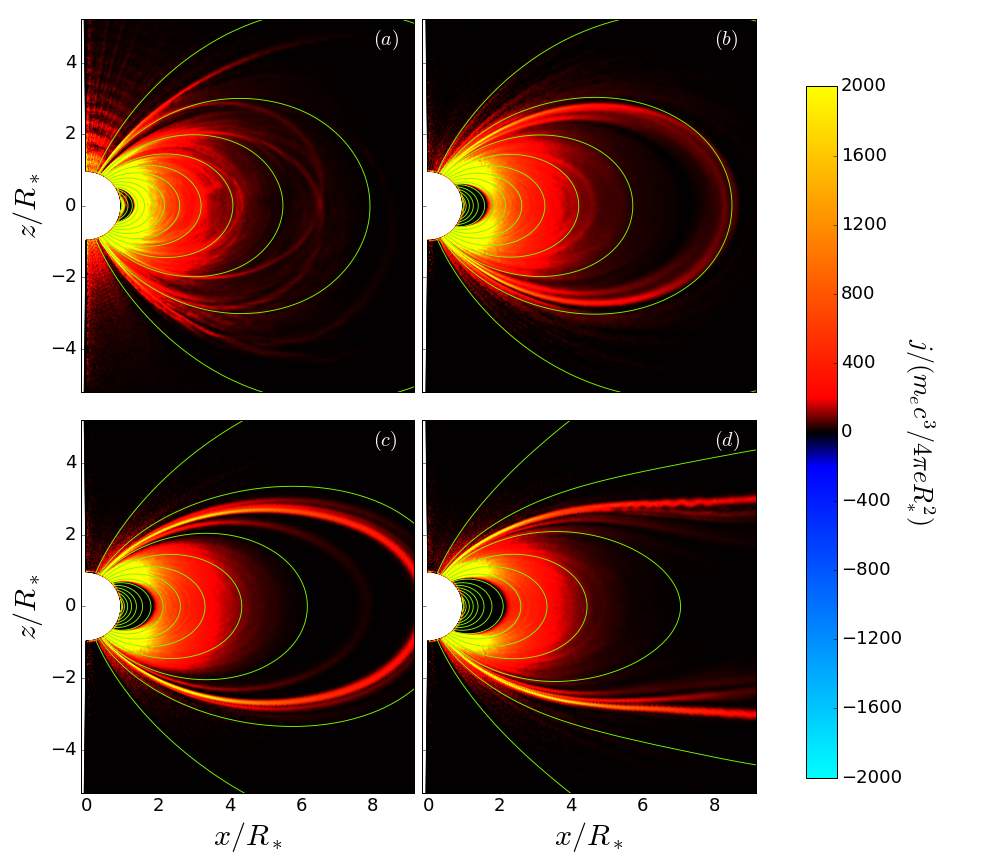}
  \caption{Color plot showing the evolution of the poloidal current density $j_p$
  in the simulation with $\gthr=10$. Four snapshots are shown: (a) $\tilde{t}=30$,
  (b) $\tilde{t}=120$,  (b) $\tilde{t}=230$, and (d) $\tilde{t}=350$.
  Note that when $j_p=0$ then also $j=0$.}
  \label{fig:current}
\end{figure*}

\begin{figure}[t]
  \centering
  \includegraphics[width=0.5\textwidth]{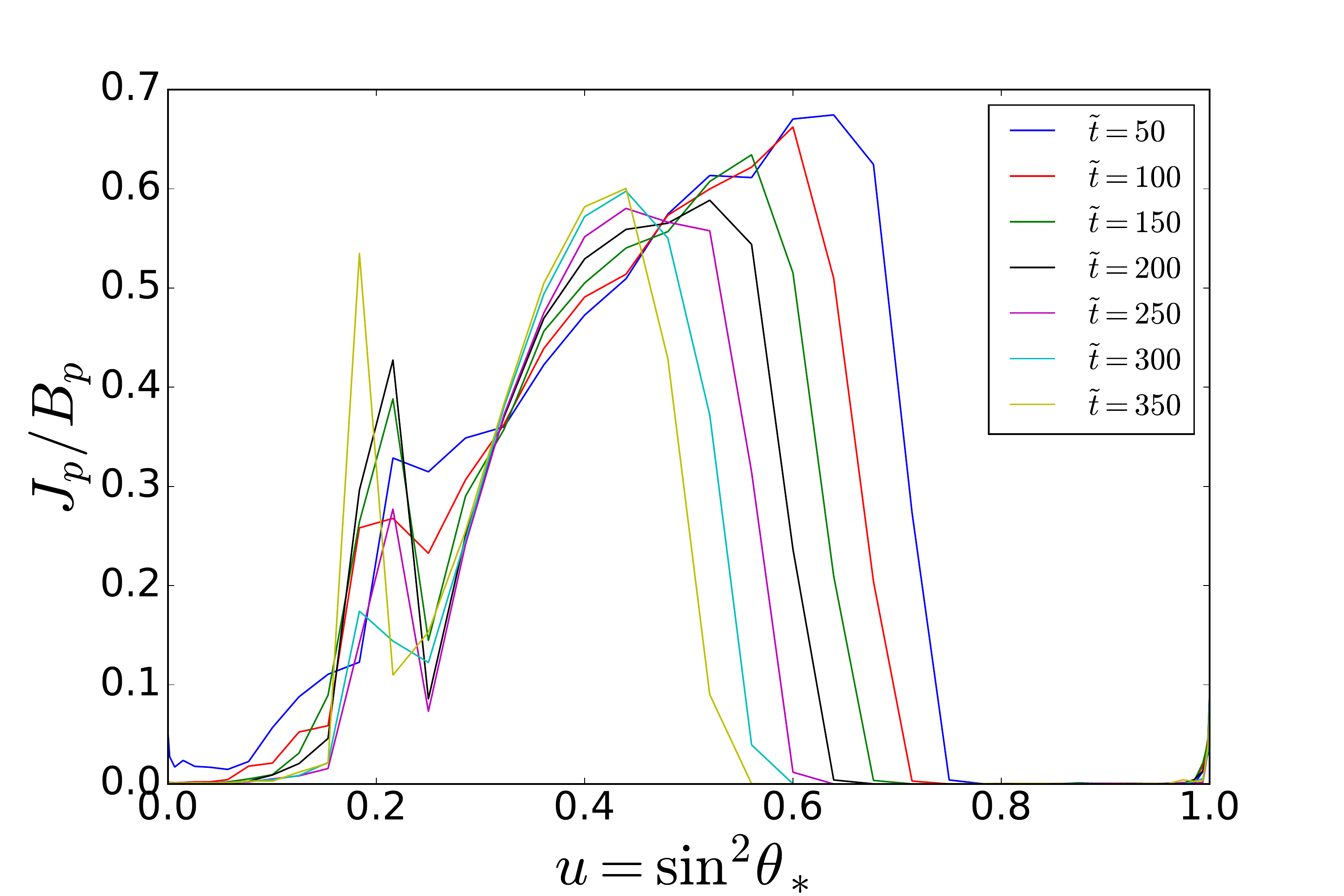}
  \caption{Evolution of the poloidal current distribution in the magnetosphere
  in the simulation with $\gthr=10$.
  The ratio $j_p/B_p$ (constant along the magnetic field lines) is shown versus
  the poloidal flux function defined in \Eq~(\ref{eq:u}); $\theta_\star$ is the polar
  angle of magnetic field line footprint on the star. The different curves show snapshots
  at times $\tilde{t}=50$, 100, 150, 200, 250, 300, and 350.}
  \label{fig:jp}
\end{figure}

Figure~\ref{fig:current} shows the resistive evolution of the j-bundle.
The untwisting of the magnetic field lines proceeds as anticipated
in
Section~\ref{sec:twist-theory}, through formation of a cavity $j=0$ that expands
from the inner magnetosphere near the equator (large flux function $u$).
Figure \ref{fig:jp} shows the evolution of the poloidal current $j_p$ until
the end of the simulation at $\tilde{t}_\mathrm{sim}=350$.
We chose to show $j_p/B_p$ because this quantity is constant along the magnetic
field lines (after averaging over short-timescale fluctuations), as expected in
a nearly force-free magnetosphere --- currents flow along the magnetic field lines.
Therefore, $j_p/B_p$ is a function of the magnetic field line, which we label
by the parameter $u=\sin^2\theta_\star$ (see \Eq~\ref{eq:u}).
Note the expansion of the region where $j_p = 0$ toward the magnetic axis,
from $u\approx 0.75$ to $u\lesssim 0.55$.

\begin{figure}[t]
  \centering
  \includegraphics[width=0.5\textwidth]{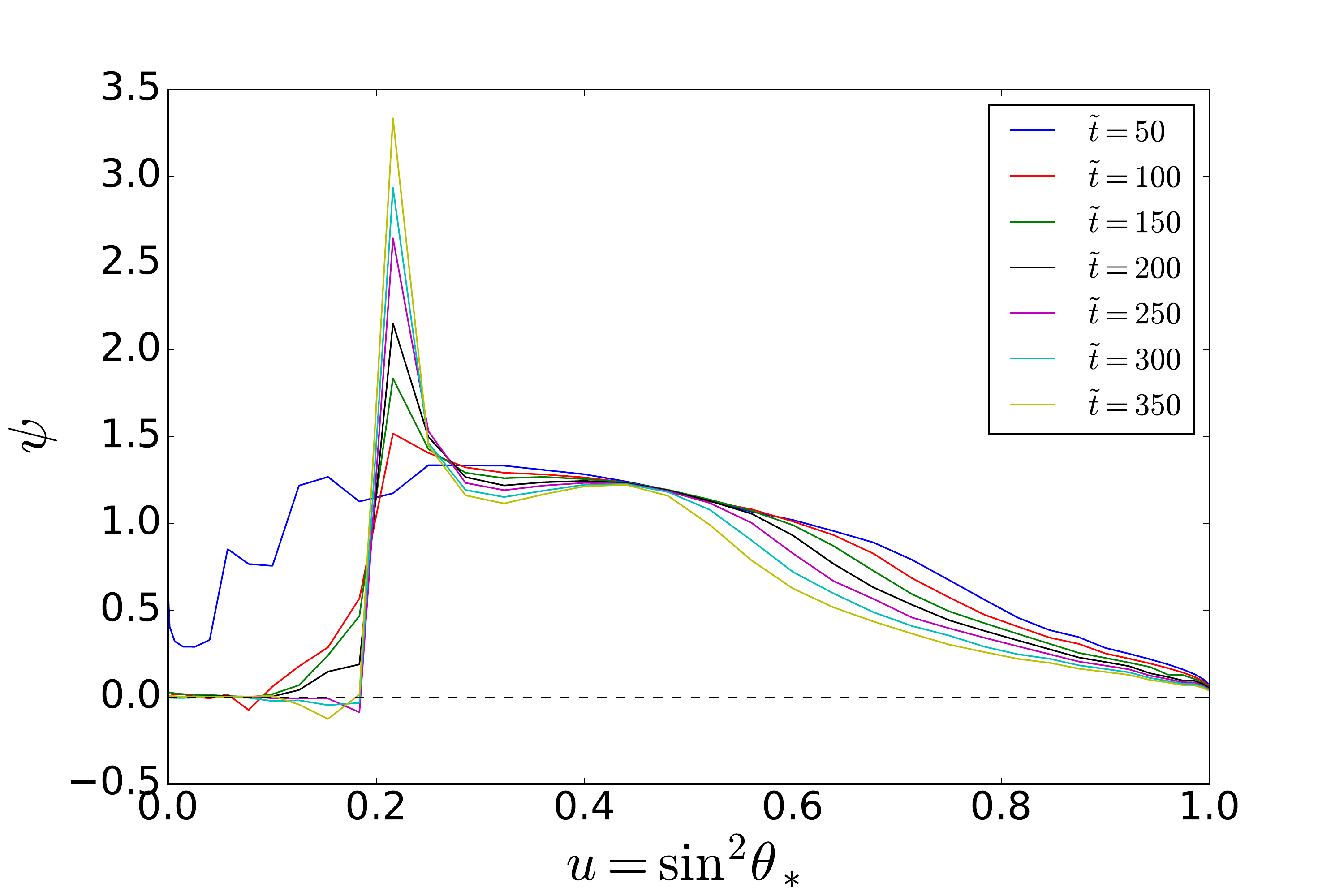}
  \caption{Evolution of the twist angle $\psi$
  in the simulation with $\gamma_\mathrm{thr} = 10$.
}
  \label{fig:twist}
\end{figure}

Figure \ref{fig:twist} shows the evolution of the integrated twist angle $\psi$
defined in Equation~\eqref{eq:twist-angle}. The untwisting proceeds from near
the equator, where the twist angle decreases over time, but the twist angle is
not simply erased, but relocated from the inner magnetosphere to the outer
parts, as expected from the untwisting Equation~\eqref{eq:twist-ev}.

A curious feature is observed to develop on the magnetic field lines with $u$
around 0.22: the twist angle $\psi$ {\it grows} and approaches 3.5 toward the end
of the simulation. This feature is also seen in the current structure shown
in Figures~\ref{fig:current} and \ref{fig:jp}. The strongly twisted, narrow bundle
of field lines is inflating with time and eventually opens up, causing a
magnetospheric instability (cf. PBH13).
Our simulation stopped right
 at the onset of this development, since we would like to limit our study to the
 quasi-steady untwisting regime.
An important difference from over-twisting studied in PBH13
is that here it is not driven by excessive surface shear. Instead, it results from
resistive evolution of the implanted twist while the crust is static.

\section{Discussion}
\label{sec:discussion}

We have performed the first axisymmetric particle-in-cell simulations of the
twisted magnetospheres of magnetars. The simulations demonstrate from
first principles that electric $e^\pm$ discharge is self-organized in the magnetosphere
to sustain the electric current $j$ demanded by the magnetospheric twist.

The results of our numerical experiment may be summarized as follows.
\begin{enumerate}
\item Shear motion of the stellar surface on a timescale $\ttw<\tev$
successfully implants a magnetic twist in the magnetosphere.
The twist is supported by continual electric current due to
self-organized $e^\pm$ discharge.

\item Particles are accelerated along the magnetic field lines to Lorentz factors
$\gamma\approx\gthr$, just sufficient to ignite pair creation.
The voltage sustaining the electric circuit, the
dissipation rate, and the lifetime of the twist are all regulated by $\gthr$.

\item Particle acceleration is localized in a gap near the equatorial plane
(Figure~\ref{fig:EdotB}).
The gap has the electric field $E_\parallel\sim 4\pi (j/c)\lambda_p$ and width
$\lgap\sim\gthr\lambda_p$, where $\lambda_p=(m_ec^3/4\pi e j)^{1/2}$ is the local
plasma skin-depth.
The plasma density in the gap is close to the minimum value $n=j/ec$ required to
conduct the electric current. Continual $e^\pm$ creation occurs near the two exits
from the gap.

\item The magnetospheric current is carried by electrons and positrons created
in the magnetosphere rather than electrons and ions extracted from the atmospheric
layer on the stellar surface. The created particles rain onto the footprints of the j-bundle,
creating two hot spots.

\item Resistive untwisting of the magnetosphere occurs on the timescale
$\tev$
estimated in \Eq~(\ref{eq:tev}), in agreement with theoretical expectations.
The evolution proceeds as predicted in B09: a cavity with $j=0$ quickly forms in
the inner magnetosphere and gradually expands, erasing the remaining electric currents.

\item A curious feature was observed in the untwisting process: while the
twist energy was decreasing as expected from ohmic dissipation,
the twist amplitude $\psi$ {\it grew} in a narrow bundle of field lines at the outer
boundary of the twisted region. This over-twisted bundle inflated so much that it
eventually opened up.
\end{enumerate}

Our results confirm that the untwisting magnetospheres naturally create shrinking hot
spots (footprints of the shrinking $j$-bundle), which have been detected in 7 transient
magnetars. The evolution timescale inferred from the simulations (\Eq~\ref{eq:tev})
is consistent with the decay timescale observed in transient magnetars (months to years).

One unknown in the setup of our numerical experiment is the profile of the
surface shear. However, basic features observed in the simulation, in particular
voltage regulation through $e^\pm$ discharge and the cavity expansion, should be
generic and independent of the details of the twist profile. It is less clear
how generic is the formation of the narrow over-twisted bundle. This could be
further explored with simulations of different shear profiles.

An important caveat in the simulation setup is the simplified ``on the spot''
prescription for pair creation, with the created $e^\pm$ pair taking a
significant energy fraction from the primary particle. As briefly discussed in
Section~\ref{sec:pairs}, this prescription is reasonable if the twist is
confined to the region of ultrastrong magnetic field near the star, $B\simgt
B_Q$. Pair creation in weaker fields tends to occur with high multiplicities,
which can launch a dense $e^\pm$ outflow and efficiently screen $E_\parallel$ in
the equatorial region \citep{2013ApJ...762...13B}. Then the gap may have to
split into two gaps and move away from the equator, closer to the star.

How the discharge will self-organize in this case can only be explored using a
more detailed implementation of the pair creation process. The future simulation
will directly track the high-energy photons produced by resonant
scattering and their conversion to pairs, without prescribing any $\gthr$.
This will be the focus of our future work, and we expect it to establish the gap location
on magnetic field lines extending far from the star. This part of the magnetosphere is
interesting for two reasons: (1) the j-bundle activity tends to
concentrate on the extended field lines, and (2) the nonthermal emission
is able to escape the outer magnetosphere while almost all resonantly scattered
photons in the region $B\gg10^{13}$~G convert to pairs \citep{2013ApJ...762...13B}.
Gap location on the extended field lines influences the hard X-ray spectrum
emitted by the twisted magnetosphere, and thus can be tested against observations.
Phase-resolved hard X-ray spectra have been measured for several magnetars
and fitted by the $e^\pm$ outflow model
\citetext{e.g. \citealp{2014ApJ...786L...1H}; \citealp{2015ApJ...807...93A}},
which assumes an electric gap near the star. Direct PIC simulations of the $e^\pm$
discharge
of high
multiplicity can verify or disprove this assumption.

We did not study in this paper what happens when the magnetosphere is
over-twisted and becomes unstable. This phenomenon is associated with the
observed giant flares of magnetars, an extreme analogy of solar flares. The
over-twisted magnetosphere inflates and creates a thin current sheet separating
magnetic fluxes of opposite polarities. The current sheet becomes
unstable to the tearing mode, which leads to magnetic reconnection and ejection
of plasmoids from the magnetosphere (\citealp{2003MNRAS.346..540L}; PBH13), resembling the
mechanism of coronal mass ejections from the sun
\cite[e.g.][]{1994ApJ...430..898M}.
Our preliminary studies using Aperture show similar behavior.
One difficulty encountered by such simulations is the huge pair creation rate
in the dissipative current sheet, which must result in quick thermalization
of the released magnetic energy. A scheme describing this transition needs to
be developed and will be a topic for future work.

\acknowledgements
This work was supported by NASA grant NNX13AI34G and
a grant from the Simons Foundation (\#446228, Andrei Beloborodov).
Some of our simulations were run on the HPC cluster Yeti at Columbia University.

\end{document}